\documentstyle[12pt]{article}

\def\Journal#1#2#3#4{{#1} {\bf #2}, #3 (#4)}

\def\NPB{{\em Nucl. Phys.} B}
\def\NPA{{\em Nucl. Phys.} A}
\def\PLB{{\em Phys. Lett.}  B}
\def\PRL{\em Phys. Rev. Lett.}
\def\PRD{{\em Phys. Rev.} D}
\def\PRC{{\em Phys. Rev.} C}

\def\be{\begin{equation}}
\def\ee{\end{equation}}
\def\bea{\begin{eqnarray}}
\def\eea{\end{eqnarray}}

\newcommand{\delslash}{\partial \hspace{-6pt}/}

\def\lnabla{\displaystyle\mathop{\nabla}^{\leftarrow}}
\def\rnabla{\displaystyle\mathop{\nabla}^{\rightarrow}}
\newcommand{\ket}{\rangle}
\newcommand{\bra}{\langle}

\begin{document}

\begin{center}
{ PHENOMENOLOGICAL STUDY OF EXCITED BARYONS~\footnote{
Talk given at the workshop {\it Future Directions 
in Quark Nuclear Physics}, Adelaide, March (1998).}}
\end{center}

\begin{center}

Atsushi Hosaka

\vspace*{0.5cm}

{\it Numazu College of Technology,
3600 Ooka,  \\ 
Numazu, Shizuoka, 410-8501  Japan\\
E-mail: hosaka@la.numazu-ct.ac.jp}\\
\end{center}


\begin{abstract}
We study baryon excited states for quark 
confinement and chiral symmetry breaking.  
In the first part we discuss spatially deformed 
baryon excitations.  
As signals of deformation, we study masses and electromagnetic 
transitions.  
Such a study of spatial structure is expected to provide 
information on quark binding mechanism and hence quark confinement.  
In the second part, we consider the 
chiral symmetry for baryons and study positive and negative parity 
baryons.  We show that there are 
two distinctive representations of chiral symmetry for baryons.  We 
investigate their phenomenological consequences in 
terms of linear sigma models.
\end{abstract}

\section{Introduction}

There are several reasons that hadron physics is one 
of interesting and hot subjects in current physics.  
In the last two decades,  much progress 
has been made to reveal rich structure of QCD from 
asymptotic to non-perturbative regions.  
Yet essential questions associated especially with non-perturbative 
properties such as 
quark confinement and chiral symmetry breaking 
are not fully understood.  
Recent developments in experiments at facilities such as 
TJNAL and COSY are important~\cite{CEBAF}.  
Using electromagnetic or hadronic probes, 
various form factors are measured in detail.  
Also a new facility SPring8 where highly polarized photon beam is 
available is expected to contribute 
to interesting hadron physics~\cite{RCNP}.  
Our objective in the present study is then a phenomenological 
study of baryon excited states, 
which are accessible by the new experiments.  
We investigate properties of baryons and attempt to 
study non-perturbative aspects of QCD.  

In the former part of this report, we discuss 
spatial structure of excited baryons.  
Previously, it was pointed out that 
observed baryon mass spectrum 
of excitation energies up to around 1 GeV  
has a structure similar to rotational band, 
which implies deformed structure for 
excited baryons~\cite{TDD,Bhaduri}.  
Recently, we have pointed out that such a deformation is a common 
property for various flavor SU(3) baryons~\cite{FET,HTT}.  
The study of spatial structure would be useful in understanding 
quark binding mechanism.  
In particular, the flavor independent nature would be an indication 
of gluon dominant dynamics.     
We study the deformed baryons in terms of a deformed oscillator 
quark (DOQ) model.  
Masses and electromagnetic transitions are investigated in detail.  

%

In the second part of this report we discuss the chiral symmetry for 
baryons.  
While the role of chiral symmetry has been extensively 
investigated for mesons, 
that for baryons is less worked out.  
For instance, it is not very well understood what
the chiral partner of the nucleon should be.  
Chiral partners are those which belong to the same multiplet of 
chiral symmetry.  
The fact that is not very well known is that 
there are two different representations of chiral 
symmetry when considering both positive and negative parity baryons.  
They lead to very different results for such as 
masses and coupling constants toward chiral 
symmetry restoration~\cite{DK}.   
Our purpose is to study the role of 
chiral symmetry for positive and negative parity 
baryons, $N$ and $N^*$. 
We study masses, $\pi NN^*$ couplings and axial charges of $N$ and 
$N^{*}$ using linear sigma models~\cite{JOH,JNOH}.

\section{Deformed oscillator quark model}

\subsection{Masses}

Let us begin with the experimental mass spectrum 
of nucleon excitations as shown in Fig. 1.  
Data are taken from the particle data group and well observed states 
with three and four stars are presented~\cite{PDG}.  
Three negative parity states of masses around 1.7 GeV and one positive 
parity state $P_{11}(1720)$ are not shown because they are out of 
systematics which we are interested in here.



From Fig.~1 we observe the followings:
\begin{enumerate}
	\item  There is degeneracy in pairs 
	($3/2^+$, $5/2^+$),
	($1/2^-$, $3/2^-$) and
	($5/2^+$, $7/2^-$).  
	This implies that the total spin $j$ of these levels are 
	composed of orbital angular momentum $l$ and spin $S = 1/2$ with 
	negligible spin-orbit forces.  

	\item  The $1/2^+$ state appears as the lowest excited states.  
	This is the Roper resonance 
	whose nature is not well understood.  

	\item  The level spacing between states of different $l$'s exhibits 
	a characteristic feature.  
	In the positive parity sector, 
	the level spacing between higher energy states is larger 
	than that of lower energy states.  
	Furthermore the level spacing of the negative parity states 
	is larger than that of those of the positive 
	parity states.  
\end{enumerate}

Here we attempt to describe these properties in terms of an
effective model.  
The third observation concerning 
the level spacing is particularly important, 
as it has lead to a model of deformed baryons whose masses appear 
as rotational band~\cite{TDD,Bhaduri}.  

In order to model the above picture, 
we consider a simple non-relativistic quark model with 
a deformed oscillator potential.  
We call this the deformed oscillator quark (DOQ) model.  
The hamiltonian of the DOQ model is then given by
\be
\label{hamiltonian}
H = \sum^3_{i=1} \left[ \frac{p_i^2}{2M_i} +
  \frac{1}{2} M_i ( \omega_x^2 x^2 + \omega_y^2 y^2
                  + \omega_z^2 z^2 )  \right] \; .
\ee
We may think that this hamiltonian
is an effective hamiltonian with the mean field 
single particle potential
which could be derived from the fundamental quark-quark interaction.  
In the DOQ model, variation of the oscillator parameter is allowed 
under the condition of volume conservation, 
$\omega_x \omega_y \omega_z = \omega_0^3$.  
We do not include more terms as meson 
degrees of freedom for an exclusive study of 
deformation effects.

The system of (\ref{hamiltonian}) 
has been worked out~\cite{Bhaduri,BM}.
After removing the center-of-mass motion,  
the intrinsic energy is given by
\be
\label{Eint}
E_{\rm int} (N_x, N_y, N_z) =
\omega_x (N_x +1) + \omega_y (N_y +1) + \omega_z (N_z +1) \, ,
\ee
where $N_i$ are the sum of oscillator 
quanta for two internal degrees of freedom, $\lambda$ and $\rho$.  
Variation of the intrinsic energy (\ref{Eint}) 
under 
$(\delta \omega_x, \delta \omega_y, \delta \omega_z )$ 
for each principal quantum number, $N = N_x + N_y + N_z$, leads 
naturally to deformed intrinsic states for $N \neq 0$.  
Energies and the corresponding shapes for several low $N$'s 
are summarized in Table 1.  

After performing the angular momentum projection~\cite{projection}, 
we obtain the rotational energy 
\be
\label{Erot}
E(N,l) = E_{\rm int}(N)
+ \frac{\hbar^2}{2I} \left[ l(l+1) - \bra l^2 \ket \right] \; 
\ee
for states of orbital angular momentum $l$. 
Here $I$ is the moment of inertia, and
$\bra l^2 \ket$ is the 
average angular momentum of the deformed intrinsic state~\cite{BM}.
Numbers for these quantities are tabulated also 
in Table~\ref{dfm_prm}.

\begin{table}[t]
	\centering
	\caption{Physical parameters of the DOQ model. \label{dfm_prm}}
	
	\begin{tabular}{ c c c c c c }
	\hline
		N & $E_{\rm int}$ $[\hbar \omega]$ & 
		$\omega^{-1}_x : \omega^{-1}_y : \omega^{-1}_z$ & shape & 
		$\hbar^2/2I$ $[\hbar \omega_0$] & $\bra l^2 \ket$  \\
	\hline
		0 & 3 & 1 : 1 : 1 &  spherical &  & 0  \\
		1 & 3.780 & 1 : 1 : 2  & prolate  & 0.126 & 3  \\
		2 & 4.327 & 1 : 1 : 3 & prolate & 0.072 & 8  \\
		\hline
	\end{tabular}	
\end{table}

We fix the constituent quark mass $M = 300$ MeV and adjust the 
absolute masses so that the nucleon $N(939)$ is reproduced.  
Taking the oscillator parameter $\omega_0 = 607$ MeV, 
we obtain theoretical mass spectrum  as shown in Fig.~1.  
The total spin $j$ is given by the coupling of the 
intrinsic spin $S=1/2$ of three quarks with the orbital
angular momentum $l$, with the spin-orbit coupling being ignored.
We find that experimental masses are nicely reproduced by the simple 
mass formula (\ref{Erot}).  
In particular, it is remarkable that 
the lowest $1/2^+$ (Roper) state is reproduced in the DOQ model.  
We have also applied this idea to flavor SU(3) 
baryons~\cite{FET,HTT} and find a good agreement between the 
theory and experiment.  
The fact that the deformed picture is commonly applied to 
various SU(3) baryons 
suggests that the underlying 
dynamics would be dominated by gluons or flavor independent 
interactions.

As a prediction of the DOQ model, we consider masses of high
spin states, e.g., $j^P = 13/2^+$ state.  
In the DOQ model, this state has the orbital angular 
momentum $l = 6$ whose energy is expected to appear around 3.5 GeV. 
This may be well compared with predictions of other models.  
For instance, the string model {\it a la}  the Regge theory predicts 
the mass of $13/2^+$ state to be around 2.5 GeV.  
The difference is substantial and should be 
studied  in future experiments.

\subsection{Electromagnetic transitions}

If spatial deformation is sufficiently developed, 
we expect evidences in various transition amplitudes.    
Here we study electromagnetic decays of excited baryons. 
Data are available in the form of helicity amplitudes~\cite{PDG}.   

Theoretically,  we study matrix elements of the 
electromagnetic interaction
\be
\label{H_em}
H_{EM} = eJ_{\mu} A^\mu \to - e \vec J \cdot \vec A \, .   
\ee
For simplicity we adopt the non-relativistic current:
\be
\vec J = \frac{1}{2m}
\left( u_f^\dagger ( i \lnabla - i \rnabla ) u_i
+
\rnabla \times (u_f^\dagger \vec \sigma u_i) 
\right) \tau_\mu \; ,
\ee
with standard notations.  
The computation of matrix elements of (\ref{H_em}) is, though 
complicated, straightforward.  
Details will be reported elsewhere~\cite{THT}.  

In Table \ref{Hamp}, 
we summarize various helicity amplitudes 
$A_{1/2}$ and $A_{3/2}$ with tentative identification of 
quark model states with physical baryon states.  
We make the following observations.  
\begin{enumerate}
	\item  In many channels, 
	theoretical predictions are roughly consistent with 
    data with small dependence on deformation.  
    The reason that the effect of the deformation is not very 
    significant is that the number of relevant degrees of freedom 
    (the valence quarks) is much less than, for example, 
    that of deformed nuclei which show clearly evidences of 
    deformation.  

	\item  The decay of the Roper 
	$N(1440)$ is not reproduced; the sign of the amplitude is wrong.  
	This sign problem has been known in the spherical quark 
	model and can not be solved in the DOQ model as well, 
	although the DOQ model works well for masses.  
	There are several possibilities which 
	could be responsible for the problem, e.g.,   
	relativistic effects~\cite{KO,Capstick} and meson cloud 
	effects~\cite{SL}.   
\end{enumerate}

As a final remark of the DOQ model, the present treatment involves a 
non-orthogonality problem.  
We have worked out the re-diagonalization of the non-orthogonal 
hamiltonian and examined its effect on masses and transition 
amplitudes~\cite{THT}.  
It turns out that the effect is not very large and the present results 
we have shown here are qualitatively unchanged.  
Therefore, we consider that the problem of the decay of the $1/2^+$ 
state is still an unsolved problem which we have to work out in 
future.

\begin{table}[t]
\caption{Helicity amplitudes in units of $GeV^{-1/2} \times 10^{-3}$.  
The column indicated as DOQ is for results of the DOQ model with 
appropriate deformation; $d=3$ for $N=2$ positive parity states 
and $d=2$ for $N=1$ negative parity states.
The column indicated as IK is for the spherical limit ($d=1$) which 
correspond to the results of Koniuk-Isgur~\cite{KI}. \label{Hamp} }
\vspace{0.2cm}
\begin{center}
\footnotesize
%
%
\begin{tabular}{ c | c c c c | c c c  c}
\hline
  Proton     &  &  & $A^{1/2}_{p}$ &  &  & $A^{3/2}_{p}$&     \\
 &  & DOQ & KI & Exp & DOQ & KI & Exp \\
\hline
1/2$^{+}$ & $^{2}S^{\prime}_{S}$ 
                          & 109  & 22.6 & -68$\pm$ 5 & --  & --  & -- & 
                          $P_{11}(1440)$\\
        & $^{2}S_{MS}$  & -14.6 & -15.9 & 
        +5$\pm$16  & --  & --  & -- & $P_{11}(1710)$\\
3/2$^{+}$ & $^{2}D_{S}$& 70.9  & 111  & 52$\pm$39  & -23.5 & -36.7& 
                                                     -35$\pm$24 & 
                                                     $P_{13}(1720)$\\
5/2$^{+}$ & $^{2}D_{S}$ & -3.8  & -5.9  & -17$\pm$10 & 47.0  & 73.5 & 
                                                         127$\pm$12 & 
                                                         $F_{15}(1680)$\\
\hline
1/2$^{-}$ & $^{2}P_{MS}$  & 151   & 156   &  
        74$\pm$11 & -- & -- & -- & $S_{11}(1535)$\\
            & $^{4}P_{MS}$  & 0     & 0     &  48$\pm$16 & -- & -- & -- & 
            $S_{11}(1650)$\\
3/2$^{-}$ & $^{2}P_{MS}$  & 24.8 & 25.6 & -23$\pm$9 & 138 & 143 & 
        163$\pm$8  & $D_{13}(1520)$\\
            & $^{4}P_{MS}$  & 0     & 0 & -22$\pm$13 & 0     & 0 & 
            0$\pm$19  & $D_{13}(1700)$ \\
5/2$^{-}$  & $^{4}P_{MS}$  & 0     & 0     & 19 $\pm$ 12  & 0     & -- 
            & 19 $\pm$ 12   & $D_{15}(1675)$ \\
\hline
  &   &     &     &   &      &  &   &  \\
\hline
 Neutron &  &  & $A^{1/2}_{n}$ &  &  & 
         $A^{3/2}_{n}$ &   \\
  & & DOQ & KI & Exp & DOQ & KI & Exp   \\
\hline
1/2$^{+}$ & $^{2}S^{\prime}_{S}$ 
                          & -73.0    & -15.1  & +39$\pm$15 & --  & --  & -- &
                          $P_{11}(1440)$\\
        & $^{2}S_{MS}$ & 4.9  & 5.3  & -5$\pm$23  & --  & --  & --  &
        $P_{11}(1710)$ \\
3/2$^{+}$ & $^{2}D_{S}$
             & -20.1 & -31.5 & -2 $\pm$26 & 0     & 0    & 
             -43$\pm$94  & $P_{13}(1720)$ \\
5/2$^{+}$ & $^{2}D_{S}$ 
           & 24.7  & 38.5  & 31$\pm$13  & 0     & 0    & -30$\pm$14 &
           $F_{15}(1680)$ \\
\hline
1/2$^{-}$ & $^{2}P_{MS}$  & -125  & -130  & 
              -72$\pm$25 & -- & -- & -- & $S_{11}(1535)$ \\
            & $^{4}P_{MS}$  & 12.9  & 13.4  & -17$\pm$37 & -- & -- & -- &
            $S_{11}(1650)$ \\
3/2$^{-}$ & $^{2}P_{MS}$  
           & -61.3  & -63.4  & -64$\pm$8 & -138  & -143  & -141$\pm$11  &
           $D_{13}(1520)$ \\
            & $^{4}P_{MS}$  & 5.8  & 6.0  & 0$\pm$56 & 30.0 & 31.0 & 
            -2$\pm$44 & $D_{13}(1700)$ \\
5/2$^{-}$    & $^{4}P_{MS}$  & -26.0  & -30.0  & -47 $\pm$ 23 & -36.8 
            & -42.4 &  -69 $\pm$ 19 & $D_{15}(1675)$ \\
\hline
\end{tabular}
\end{center}
\end{table}

\section{Chiral symmetry of baryons}

In the latter part of this report, we discuss chiral symmetry of 
baryons~\cite{JOH,JNOH}.  
Although chiral symmetry and its spontaneous breakdown 
play important role for low energy hadron physics, its actual role 
for baryon dynamics is not well understood.  
For example, when chiral symmetry is restored, 
we expect the appearance of degenerate particles which belong to the 
same multiplet of the chiral group (chiral partners).  
Since these particles have opposite 
parities, one might expect that the $N(939) \equiv N$ 
and $N(1535) \equiv N^*$ might be candidates of such chiral partners.  
If this would be the case, it is an interesting question what 
properties of $N$ and $N^*$ are governed by chiral symmetry.  
In the following we consider the chiral $SU(2)_L \times SU(2)_R$ 
group.  

The fact that is not very well known is  
that when we consider two kinds of baryons, there
are two distinctive representations of the chiral group.   
One is called the naive and the other mirror representation~\cite{DK}.  
It turns out that they lead to very 
different results for properties such as the mass, meson-baryon 
couplings and axial charges.  

Let us define the naive nucleons ($N_{1}, N_{2}$) and the mirror nucleons 
($\psi_{1}, \psi_{2}$) such that they transform 
under $SU(2)_L \times SU(2)_R$ chiral transformations as
\bea
  N_{1R} \longrightarrow R N_{1R}\  & & \hspace{1cm}  N_{1L}
  \longrightarrow L N_{1L} \ , \nonumber \\
  N_{2R} \longrightarrow R N_{2R}\  & & \hspace{1cm}  N_{2L}
  \longrightarrow L N_{2L} \ . \label{Tr_naive}
\eea
and 
\bea
   \psi_{1R} \longrightarrow R \psi_{1R} \  & , \hspace{1cm} &   \psi_{1L}
   \longrightarrow L \psi_{1L} \ ,  \nonumber \\
   \psi_{2R} \longrightarrow L \psi_{2R} \  & , \hspace{1cm} &   \psi_{2L}
   \longrightarrow R \psi_{2L} \ . \label{Tr_mirror}
\eea
In the naive representation the left and right components transform in 
the same way, while in the mirror representation they transform in 
the opposite way.  
The reason that the latter is possible is 
that while the chirality of the fermion is associated with 
the Lorentz group, that of chiral symmetry is 
associated with internal symmetry.  
Therefore, the left and 
right handed fermions can take both representations of the internal 
chiral group.

Now we construct linear sigma models which respect chiral 
symmetry.  
Knowing the transformation (\ref{Tr_naive}) and (\ref{Tr_mirror}) 
together with that of the meson field, the chiral invariant 
lagrangians for the nucleon sector 
can be written up to fourth order in mass dimension: 
\bea
{\cal L}_{\rm{naive}} & = &\bar{N_1} \delslash N_1 + \bar{N_2}
\delslash
   N_2 + a \bar{N_1} (\sigma + i \gamma_5 \vec{\tau} \cdot \vec{\pi})
   N_1 + b \bar{N_2} (\sigma + i \gamma_5 \vec{\tau} \cdot \vec{\pi})
   N_2 \nonumber \\
         & & + c \{ \bar{N_2} (\gamma_5 \sigma + i \vec{\tau} \cdot
         \vec{\pi}) N_1  - \bar{N_1} (\gamma_5 \sigma + i \vec{\tau} \cdot
         \vec{\pi}) N_2 \} + {\cal L}_{M}  \, , 
\label{L_naive} \\
{\cal L}_{\rm{mirror}} & = & \bar{\psi_1} i \delslash \psi_1 +
           \bar{\psi_2} i \delslash \psi_2
        + m_{0}( \bar{\psi_2} \gamma_{5} \psi_1 - \bar{\psi_1}
            \gamma_{5} \psi_2  )
                \nonumber \\
      & &    + a \bar{\psi_{1}} (\sigma + i \gamma_5 \vec{\tau} \cdot
           \vec{\pi}) \psi_{1} +
        b \bar{\psi_{2}} (\sigma - i \gamma_5 \vec{\tau} \cdot
        \vec{\pi}) \psi_{2} \, .
\label{L_mirror}
\eea
Here $a$, $b$ and $c$ are coupling constants.  
What should be emphasized here is that 
in the mirror Lagrangian, the chiral invariant mass 
term is allowed with the mass parameter $m_0$.  
Thus both $\psi_1$ and $\psi_2$ remain massive with a degenerate mass 
when chiral symmetry is restored (parity doubling of 
baryons)~\cite{DK}.

%
%

In the lagrangians (\ref{L_naive}) and (\ref{L_mirror}), there is a 
mixing term between the two nucleons, when chiral symmetry is 
spontaneously broken ($\sigma \to \sigma_0$).  
The non-diagonal mass terms can be diagonalized by the physical 
fields, and the resulting eigenvalues are given by
\bea
   m_{\pm} = {\sigma_{0} \over 2} \left( \sqrt{(a + b)^{2} + 4c^{2}}
   \mp (a-b) \right) \, , \; \; \; \; &\;& \; \; \; \; 
   {\rm Naive}  \label{naimass} \\
  m_\pm = {1 \over 2} ( \sqrt{ (a+b)^2 \sigma_0^2 + 4 m_0^2}
              \mp (a -b) \sigma_0 ) \, \; \; \; \; &\;& \; \; \; \; 
   {\rm Mirror}  \label{mirmass}
\eea
A schematic plot of the masses are shown in Fig~2.  
In the naive case, positive and negative parity nucleons 
degenerate with the vanishing mass when chiral symmetry is restored, 
while they have the finite mass 
$m_0$ in the mirror case.

We can also study other physical quantities.  
Here we just quote a few results of interest.  
\begin{enumerate}
	\item  The $\pi NN^*$ coupling constant vanishes in the naive case, 
	while it can be finite in the mirror case.  
	The discrepancies in the previous QCD sum rule 
	analyses~\cite{jko,kimlee} can be 
	explained by the different choices of the interpolating field 
	corresponding to either naive or mirror representation.  

	\item  Chiral multiplets are different for the naive and mirror 
	models.   
	In the naive case,  positive and negative parity nucleons belong to 
	separate multiplets, while they do in the same multiplet in the 
	mirror case~\cite{JOH}.  

	\item  The relative sign of the axial charges of the positive and 
	negative parity nucleons is different.  
\end{enumerate}
These properties are summarized in Table~\ref{sum}.  
They can be used as signals to distinguish the chiral representations 
of baryons.  
In particular the behaviors of coupling constants~\cite{nemoto} 
and axial charges are 
interesting.  
As pointed out in recent report~\cite{JKimO}, density dependence of 
the $\pi NN^*$ coupling is also a useful signal.  
Experimentally, we are able to study these quantities associated 
with $N(1535)$ (assuming that $N(1535)$ is the chiral partner of the 
nucleon) through productions of an $\eta$ meson.  
Such processes have been studied, but not much emphasis has been 
put on chiral symmetry of baryons.  
Certainly chiral symmetry of baryons is an interesting topic in 
both theory and experiment.

\begin{table}
\caption{Comparison of the naive and mirror constructions\label{sum}}
\begin{center}
\begin{tabular}{c c c}
\hline
      & naive  & mirror \\
\hline
definition & $N_{2R} \rightarrow R N_{2R}$ & $ \psi_{2R} 
           \rightarrow L \psi_{2R}$\\
           & $N_{2L} \rightarrow L N_{2L} $ & $\psi_{2L} 
           \rightarrow R \psi_{2L} $\\
mass in the & & \\
symmetric phase& 0 & $m_0$ (finite) \\
$\pi NN^* $ coupling & 0 & $ (a + b)/\cosh \delta$ \\
chiral partner & $N_+ \leftrightarrow \gamma_5 N_+ \ , \
                  N_- \leftrightarrow \gamma_5 N_- $ &
       $\psi_+ \leftrightarrow \psi_- $ \\
$g_{A}^{NN} g_{A}^{N^* N^*} $ & positive & negative  \\
\hline
\end{tabular}
\end{center}
\end{table}


\section{Summary}

We have overviewed  baryon excited states based on two different 
points of view. 

Low lying mass spectrum up to about 1 GeV suggests an interesting 
possibility of deformed baryons.  
A simple DOQ model has  been quite useful in reproducing the mass 
spectrum including the Roper state which has been difficult to 
be reproduced in conventional quark models.  
Such properties should be further studied through various form factors.  
Not only electromagnetic decays, a systematic study 
of strong decays should also be useful.  

We have also emphasized the role of chiral symmetry for baryons 
and pointed out that there are 
two distinctive realizations of chiral symmetry.  
They predict very different results for baryon properties, and 
we need more works in understanding the role of chiral symmetry for 
baryons.   

So far we have discussed the two aspects of baryons, spatial structure which 
is perhaps related to quark confinement, and chiral symmetry, in an 
independent way.  
We do not yet know their link from more fundamental point of view.  
We need more microscopic theories which can describe 
both quark confinement and 
chiral symmetry breaking.  
For this, we should mention that 
there is a promising theory such as the DGL theory~\cite{DGL}.  
In any event, baryon physics is an exciting subject from which 
we expect to learn more on non-perturbative nature of QCD.

\section*{Acknowledgments}
The former part of this report is based on the collaboration with 
Hiroshi Toki and Miho Takayama at RCNP (Osaka).  
Materials in the latter part stem from the collaboration with Daisuke 
Jido, Yukio Nemoto and Makoto Oka at Tokyo Institute of Technology.  
The author would like to thank all of them for stimulating 
discussions.   
He also thanks the members at the Center for the Subatomic Structure of 
Matter at Adelaide for kind hospitality during the workshop 
on {\it Future Directions 
in Quark Nuclear Physics}, March (1998).  
This work is supported in part by Grant-in-Aid (09740224) 
from Ministry of Education.


\newpage

\section*{Figure Captions}

\begin{description}
	\item[Fig. 1]  
	Observed masses of nucleon excitations as compared with the 
    results of the DOQ model.

	\item[Fif. 2]  
	A schematic plot of masses of  
	$N$ and $N^*$ as functions of $\sigma_0$.

\end{description}


\begin{thebibliography}{99}


 \bibitem{CEBAF}  See for example, Proceedings of 
 CEBAF/INT workshop on $N^*$ physics, Seattle, Sep.(1996).

 \bibitem{RCNP}  Workshop on future projects in RCNP, Osaka, March 
 (1998).  

 \bibitem{TDD}  H. Toki, J. Dey and M. Dey, 
  \Journal{\PLB} {133}{20}{1983}.

 \bibitem{Bhaduri}
     R.K. Bhaduri, B.K. Jennings and J.C. Wadington, 
     \Journal{PRD} {29}{2051}{1984};
     M.V.N. Murthy e{\it et al}, 
     \Journal{PRD} {30}{152}{1984};
     M.V.N. Murthy {\it et al},  
     \Journal{ZPC} {29}{385}{1985}.

  \bibitem{FET}  
  H. Fujimura, H. Ejiri and H. Toki, Proceedings of {\it Frontier96}, 
  RCNP, Osaka (1996).  

  \bibitem{HTT}  
  A. Hosaka, H. Toki and M. Takayama, {\em Int. J. Mod. Phys. Lett.} A 
  13 (1998) 1699.   

   \bibitem{DK} C. DeTar and T. Kunihiro,
    \Journal{\PRD}{39}{2805}{1989}.

  \bibitem{JOH} D. Jido, M. Oka and A. Hosaka, 
  \Journal{\PRL} {80}{448}{1988}. 

  \bibitem{JNOH} D. Jido, Y. Nemoto M. Oka and A. Hosaka, 
  TIT preprint (1988), hep-ph/9805306.  

 \bibitem{PDG} R.M. Barnett et al. (Particle Data Group), 
 \Journal{\PRD} {54}{1}{1996}. 

  \bibitem{BM} 
  A. Bohr and B. Mottelson, "Nuclear Structure", Benjamin Inc. (1975).

  \bibitem{projection}
  R.E. Peierls and J. Yoccoz, {\em Proc. Phys. Soc. London} 70 381 (1957).

  \bibitem{THT} M. Takayama, H. Hosaka and H. Toki, in preparation 
  (1998).  

  \bibitem{KI} R. Koniuk and N. Isgur, 
  \Journal{\PRD} {21}{1868}{1980}.

  \bibitem{KO} 
  T. Kubota and K. Ohta, 
  \Journal{\PLB} {65}{374}{1976}.  

  \bibitem{Capstick}
  S. Capstick, \Journal{\PRD} {46}{1965}{1992}; {\it ibid.} 2864.  

  \bibitem{SL}
  T. Sato and T.-S.H. Lee, 
  \Journal{\PRC} {54}{2660}{1996}.  


    \bibitem{jko} D. Jido, N. Kodama and M. Oka,
    \Journal{\PRD}{54}{4532}{1996};
    D. Jido and M. Oka,  TIT preprint, hep-ph/9611322.

    \bibitem{kimlee} H. Kim and S.-H. Lee,
    \Journal{\PRD}{56}{4278}{1997};
     S.-H. Lee and H. Kim, 
    \Journal{\NPA}{612}{418}{1997}.  


    \bibitem{nemoto}  Y. Nemoto, D. Jido, M. Oka and A. Hosaka,
    \Journal{\PRD}{57}{4124}{1998}
    
    \bibitem{JKimO}  H. Kim, D. Jido and M. Oka, TIT preprint (1998), 
    hep-ph/9806275.  

    \bibitem{DGL}
    H. Toki, H. Suganuma and S. Sasaki, 
    \Journal{\NPA}{577}{353c}{1994};
    H. Suganuma, S. Sasaki and H. Toki, 
    \Journal{\NPB}{435}{207}{1995}. 

\end{thebibliography}
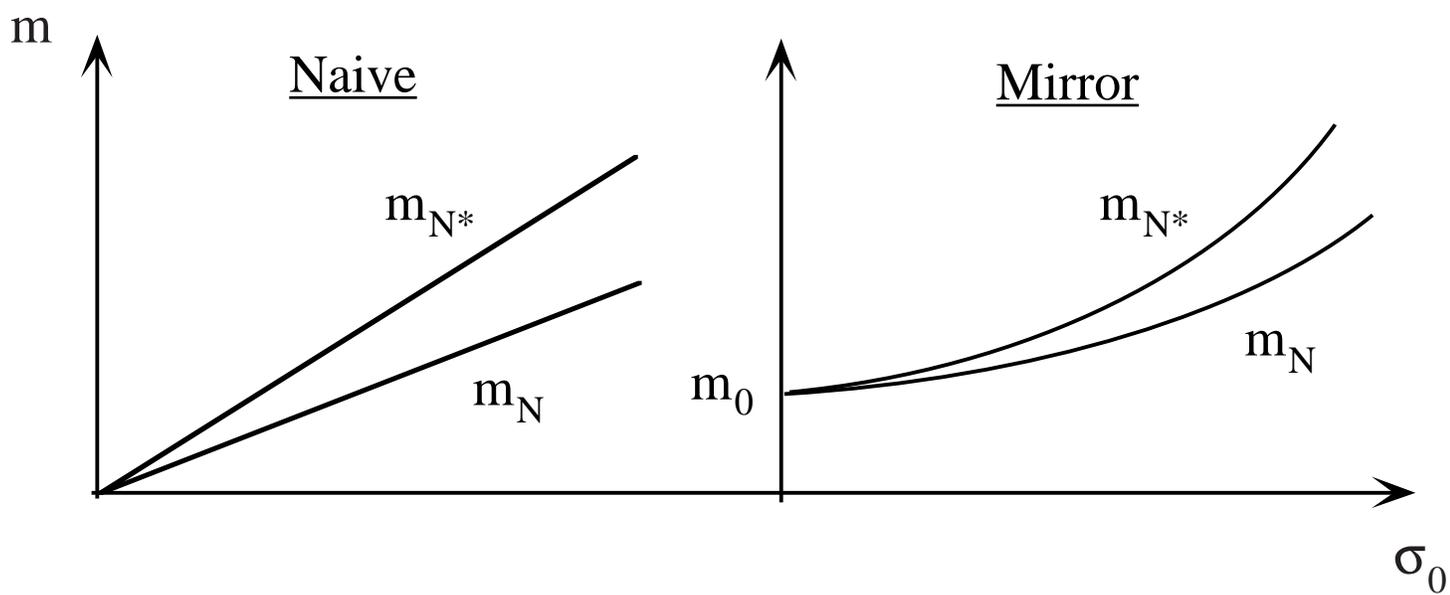
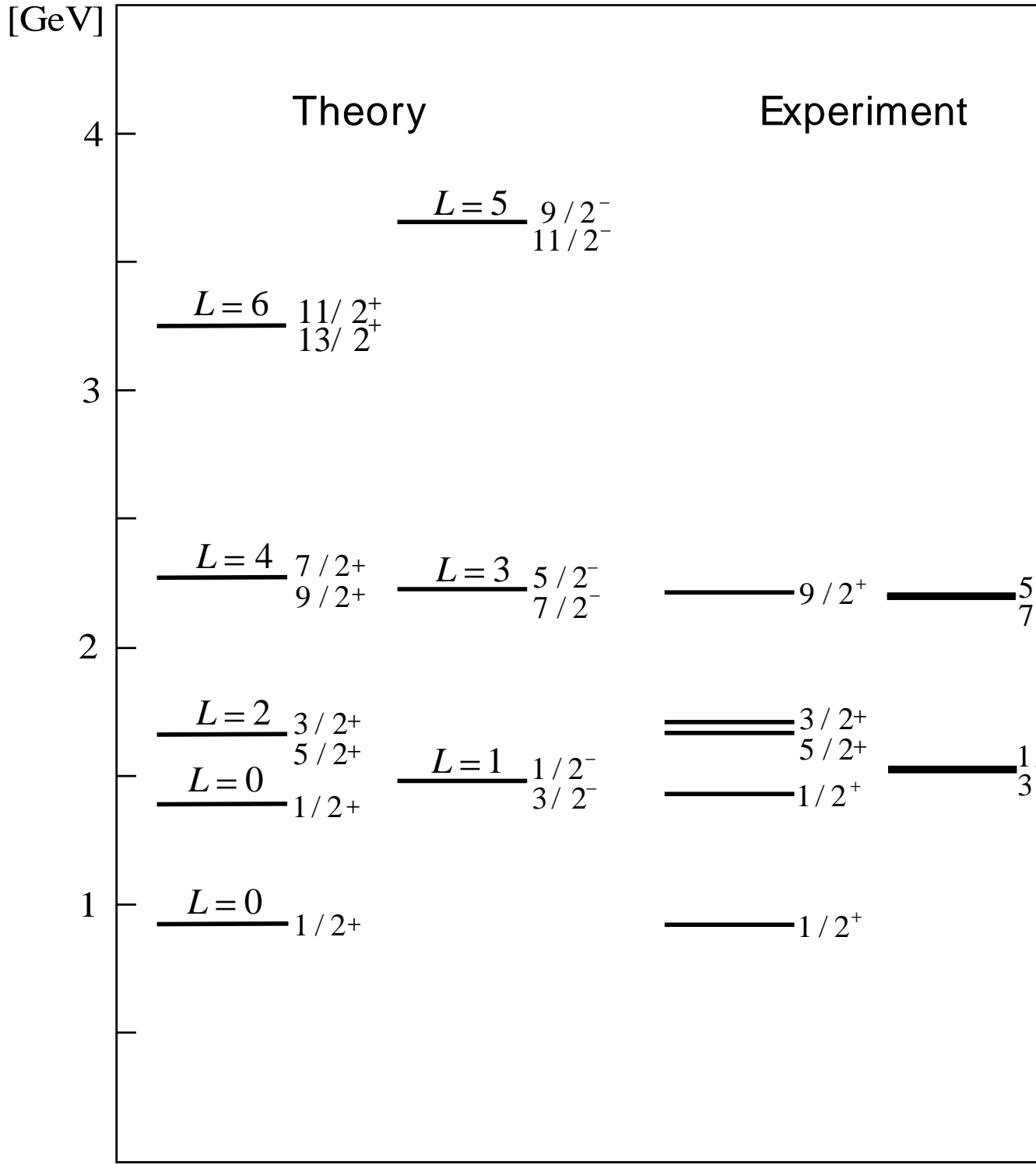
\end{document}